\documentstyle[11pt,aaspp4]{article}

\begin{document}

\title{A Preliminary Visual Orbit of BY~Draconis}
\author{A.F.~Boden\altaffilmark{1,2},
	B.F.~Lane\altaffilmark{3,2}
}
\altaffiltext{1}{Infrared Processing and Analysis Center, California Institute of Technology}
\altaffiltext{2}{Jet Propulsion Laboratory, California Institute of Technology}
\altaffiltext{3}{Geology and Planetary Sciences, California Institute of Technology}

\authoremail{bode@ipac.caltech.edu}

\begin{abstract}
We report on the preliminary determination of the visual orbit of the
double-lined spectroscopic binary system BY~Draconis with data
obtained by the Palomar Testbed Interferometer in 1999.  BY~Dra is a
nearly equal-mass double-lined binary system whose spectroscopic orbit
is well known.  We have estimated the visual orbit of BY~Dra from our
interferometric visibility data fit both separately and in conjunction
with archival radial velocity data.  Our BY~Dra orbit is in good
agreement with the spectroscopic results.  Due to the orbit's face-on
orientation, the physical parameters implied by a combined fit to our
visibility data and radial velocity data do not yet result in precise
component masses and a system distance, but with continued
interferometric monitoring we hope to improve the mass estimates to
better than 10\% determinations.

\end{abstract}

\section{Introduction}

BY~Draconis (HDE 234677, Gl 719) is a well-studied nearby ($\sim$ 15
pc), multiple stellar system containing at least three objects.  The A
and B components form a short-period (6 d) late-type (K6~Ve --
K7~Vvar) binary system, whose spectroscopic orbit is well known
(\cite{Bopp73}, hereafter BE73; \cite{Vogt79}, hereafter VF79; and
\cite{Lucke80}, hereafter LM80).  BY Dra is the prototype of a class
of late-type flare stars characterized by photometric variability due
to star spots, rapid rotation, and Ca II H and K emission lines.  Like
the BY~Dra system itself, a large fraction ($>$ 85\%,
\cite{Bopp77,Bopp80}) of BY Dra stars are known to be in short-period
binary orbits.  The rapid rotation of BY Dra A (period 3.83 d) that
gives rise to the spotting and photometric variability is consistent
with pseudosynchronus rotation with the A -- B orbital motion
(\cite{Hut81,Hall86}), but pseudosynchronization is disputed by
Glebocki \& Stawikowski (1995, 1997) who assert asynchronous rotation
and roughly 30$^\circ$ misalignment of the orbital/rotational angular
momentum vectors.

Despite the fact that the BY~Dra A and B components are nearly equal
mass, the system exhibits a significant brightness asymmetry in
spectroscopic studies (VF79, LM80).  VF79 attributes this to the
hypothesis that A and B components of BY Dra are pre-main sequence
objects (VF79, \cite{Bopp80}), and are still in the contraction phase.
VF79 argues for physical sizes of the A component in the range of 0.9
-- 1.4 R$_{\sun}$, based primarily on rotation period and $v \sin i$
considerations.  LM80 concur with the A component physical size
argument from their $v \sin i$ measurements, estimating a 1.2
R$_{\sun}$ size for a $\sin i \approx$ 0.5 (presuming rotation/orbit
spin alignment with pseudosynchronization, and our orbital inclination
from Table \ref{tab:orbit}).  However, they continue by pointing out
that if the A component macroturbulance were significantly larger than
solar, then the $v \sin i$ measurements and the component diameters
they are based on are biased high.  If the pre-main sequence
interpretation is correct, the BY~Dra components are additionally
interesting as an examples of the transition region between pre and
zero-age main sequence states.

BY Dra was detected as a hierachical triple system through common
proper motion measurements of a BY Dra C component by Zuckerman et
al.~(1997); they find the C component separated by 17'' from the A --
B pair.  Zuckerman's photometry on the C component is consistent with
an M5 main-sequence interpretation ($V$ - $K$ $\approx$ 6.2); assuming
all three stars are coeval this clearly poses problems for the VF79
pre-main sequence hypothesis for the A and B components.  At a
projected physical separation of approximately 260 AU from the A -- B
binary, the putative low-mass C component would have negligible
dynamical influence on the A -- B binary motion.  Further, the
Hipparcos catalog (\cite{HIP97}) implies the presence of at least one
additional component as it lists BY Dra as having a circular
photocentric orbital solution with 114 d period.  This 114 d period is
previously unreported in spectroscopic studies, and if correct it is
difficult to understand why this motion was not previously detected.

Herein we report on a preliminary determination of the BY Dra A -- B
system visual orbit from near-infrared, long-baseline interferometric
measurements taken with the Palomar Testbed Interferometer (PTI).  PTI
is a 110-m $H$ (1.6$\mu$m) and $K$-band (2.2 $\mu$m) interferometer
located at Palomar Observatory, and described in detail elsewhere
(\cite{Colavita99a}).  PTI has a minimum fringe spacing of roughly 4
milliarcseconds (10$^{-3}$ arcseconds, mas) in $K$-band at the sky
position of BY~Dra, allowing resolution of the A -- B binary system.

\section{Observations}
\label{sec:observations}
The interferometric observable used for these measurements is the
fringe contrast or {\em visibility} (squared, $V^2$) of an observed
brightness distribution on the sky.  The reader is refered to one of
the previous papers in this series
(\cite{Boden99a,Boden99b,Boden2000}) for descriptions of the $V^2$
observables for binary stars.

BY~Dra was observed in conjunction with objects in our calibrator list
(Table \ref{tab:calibrators}) by PTI in $K$-band ($\lambda \sim 2.2
\mu$m) on 21 nights between 23 June 1999 and 21 October 1999, covering
roughly 20 periods of the system.  Additionally, BY~Dra was observed
by PTI in $H$-band ($\lambda \sim 1.6 \mu$m) on 13 September, 2
October, and 20 October 1999.  BY~Dra, along with calibration objects,
was observed multiple times during each of these nights, and each
observation, or scan, was approximately 130 sec long.  For each scan
we computed a mean $V^2$ value from the scan data, and the error in
the $V^2$ estimate from the rms internal scatter.  BY~Dra was always
observed in combination with one or more calibration sources within
$\sim$ 10$^{\circ}$ on the sky.  Table \ref{tab:calibrators} lists the
relevant physical parameters for the calibration objects.  We have
calibrated the $V^2$ data by methods discussed in Boden et al.~(1998).
Our 1999 observations of BY~Dra result in 67 calibrated visibility
measurements (50 in $K$-band, 17 in $H$-band).  One notable aspect of
our BY~Dra observations is that its high declination (51$^\circ$)
relative to our Palomar site (33$^\circ$ latitude) puts it at the
extreme Northern edge of the delay line range on our N-S baseline
(\cite{Colavita99a}), implying extremely limited $u-v$ coverage on
BY~Dra.  To our PTI visibilities we have added 44 double-lined radial
velocity measurments: 14 from BE73, seven from VF79, and 23 CORAVEL
measurements from LM80.

\begin{table}[t]
\begin{center}
\begin{small}
\begin{tabular}{ccccc}
\hline
Object    & Spectral & Star        & BY~Dra         & Adopted Model \\
Name      & Type     & Magnitude   & Separation     & Diameter (mas)  \\
\hline
HD 177196 & A7 V     & 5.0 V/4.5 K & 6.6$^{\circ}$  & 0.70 $\pm$ 0.06   \\
HD 185395 & F4 V     & 4.5 V/3.5 K & 9.9$^{\circ}$  & 0.84 $\pm$ 0.04   \\
\hline
\end{tabular}
\caption{PTI BY~Dra Calibration Objects Considered in our Analysis.
The relevant parameters for our two calibration objects are
summarized.  The apparent diameter values are determined from
effective temperature and bolometric flux estimates based on archival
broad-band photometry, and visibility measurements with PTI.
\label{tab:calibrators}}
\end{small}
\end{center}
\end{table}

\section{Orbit Determination}

As in previous papers in this series
(\cite{Boden99a,Boden99b,Boden2000}) the estimation of the BY~Dra
visual orbit is made by fitting a Keplerian orbit model directly to
the calibrated (narrow-band and synthetic wide-band) $V^2$ and RV data
on BY~Dra; because of the limited $u-v$ coverage in our data
derivation of intermediate separation vector models is impossible.  The
fit is non-linear in the Keplerian orbital elements, and is therefore
performed by non-linear least-squares methods with a parallel
exhaustive search strategy to determine the global minimum in the
chi-squared manifold.  The reader is refered to the previous papers
for futher details on our orbit estimation procedures.

\begin{figure}
\epsscale{0.7}
\plotone{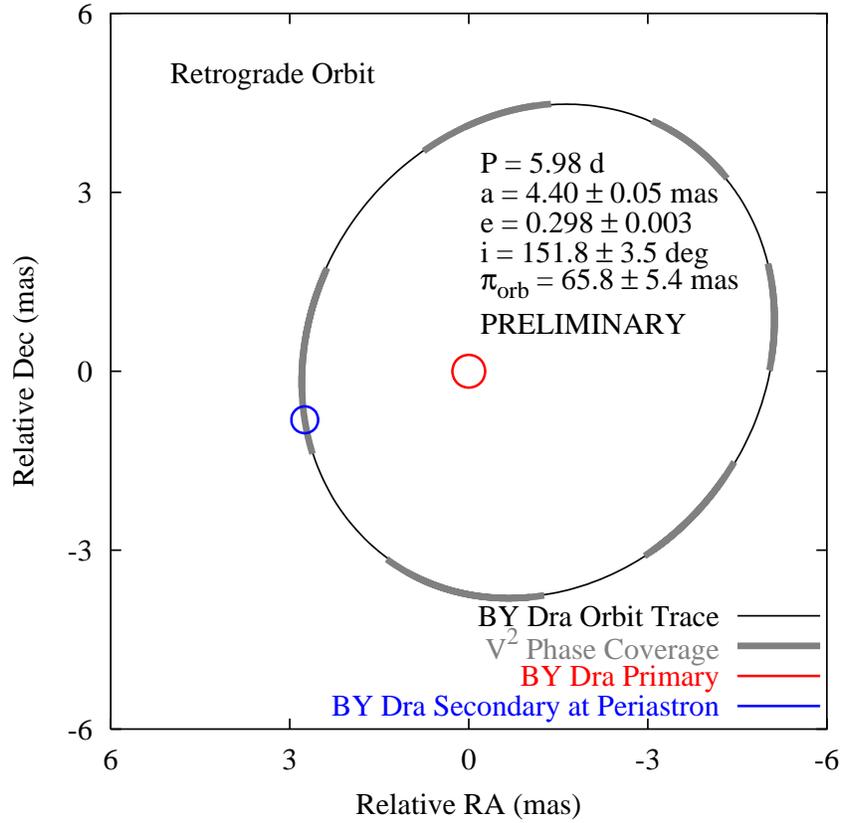}
\caption{Visual Orbit of BY~Dra.  The relative visual orbit model of
BY~Dra is shown, with the primary and secondary objects rendered at
T$_0$ (periastron).  The heavy lines along the relative orbit indicate
areas where we have orbital phase coverage in our PTI data (they are
not separation vector estimates); our data samples most phases of the
orbit well, leading to a reliable orbit determination.  Component
diameter values are estimated (see discussion in
\S~\ref{sec:discussion}), and are rendered to scale.
\label{fig:BYDra_orbit}}
\end{figure}

Figure \ref{fig:BYDra_orbit} depicts the relative visual orbit of the
BY~Dra system, with the primary component rendered at the origin, and
the secondary component rendered at periastron.  We have indicated the
phase coverage of our $V^2$ data on the relative orbit with heavy
lines; our data samples most phases of the orbit well, leading to a
reliable orbit determination.  The apparent inclination is very near
the estimate given by VF79 based on the primary rotation period,
assumed size, and assumption of parallel orbital/rotational angular
momentum alignment.  The orbit is seen approximately 30$^\circ$ from a
face-on perspective, which makes physical parameter determination
difficult (\S \ref{sec:discussion}).

\begin{figure}
\epsscale{0.8}
\plotone{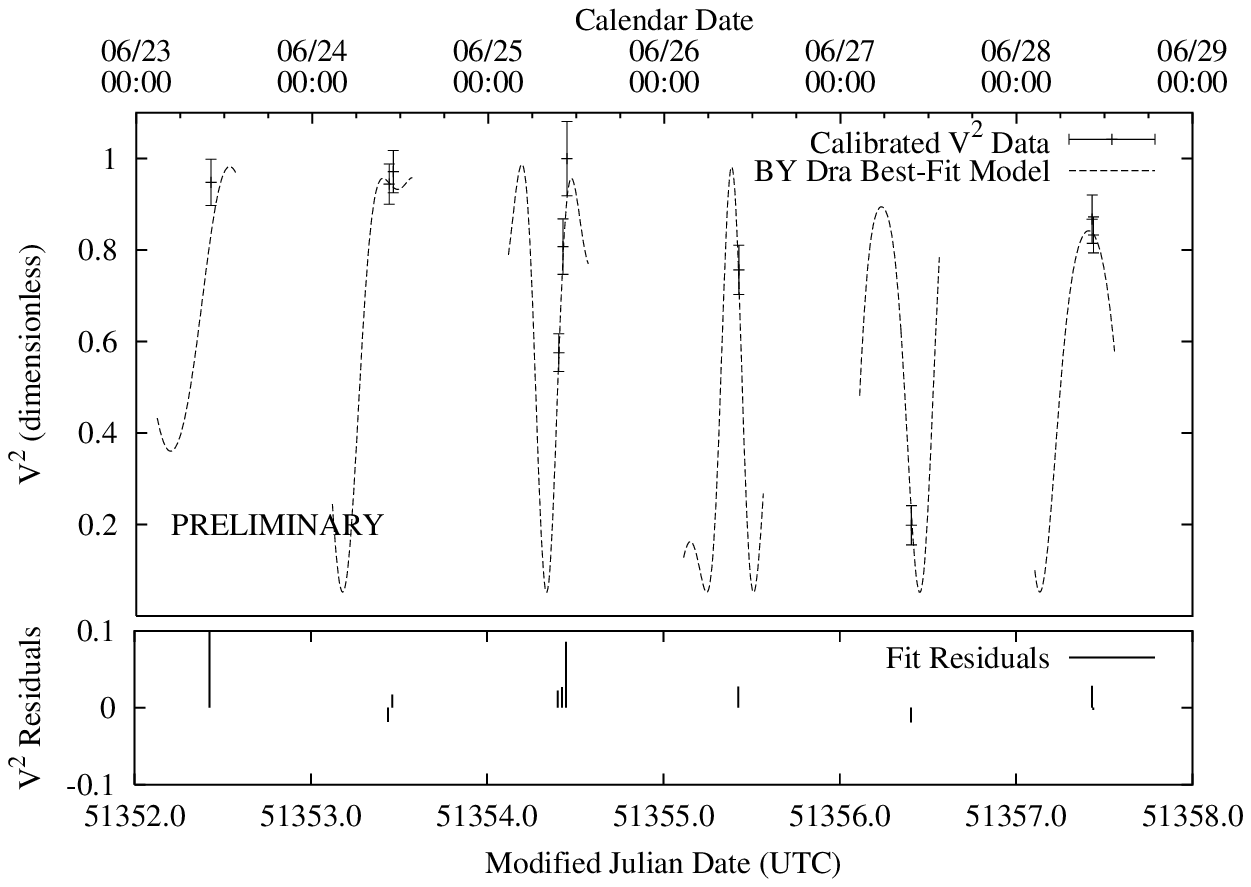}\\
\plotone{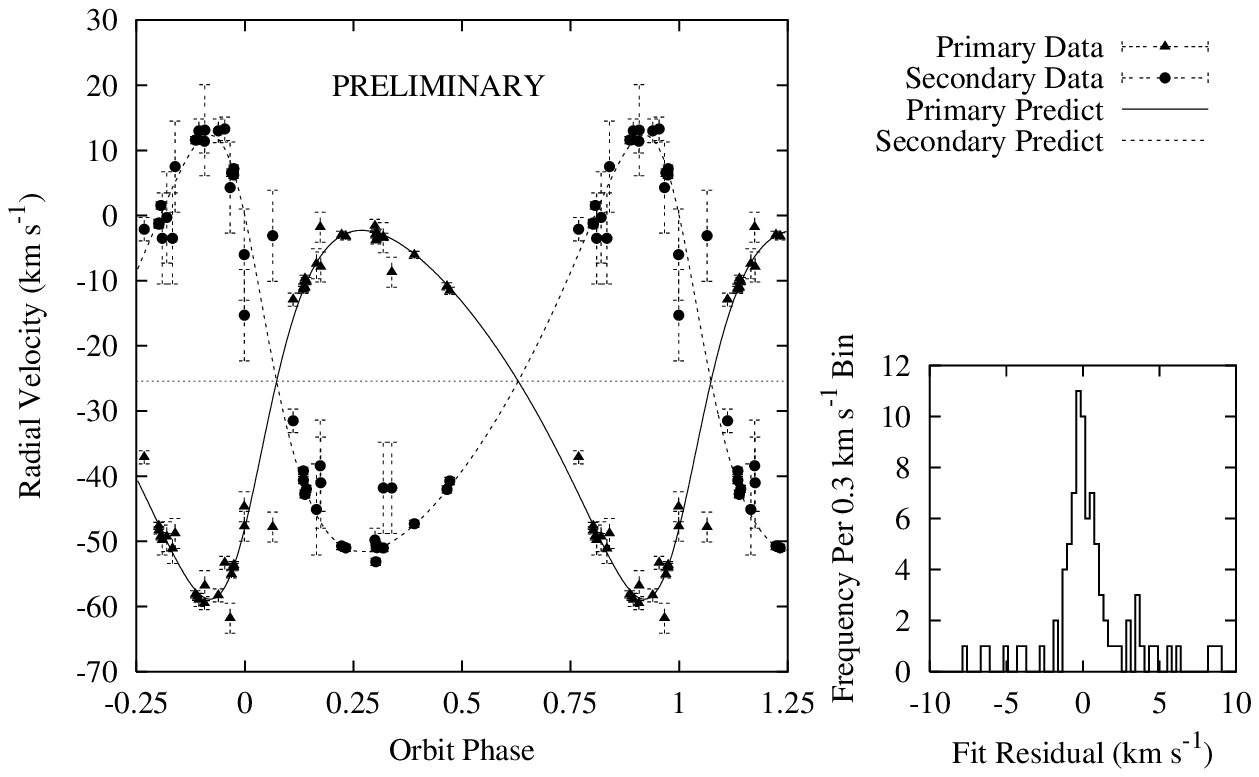}
\caption{Joint $V^2$/Radial Velocity Fit of BY~Dra.  a: Six
consecutive nights of $V^2$ data on BY~Dra, and best-fit model
predictions.  b: Phased archival RV data (BE73, VF79, LM80) and RV
predictions from our best-fit orbital model.  Inset is a histogram of
RV residuals to the fit model.
\label{fig:BYDraFit}}
\end{figure}

Figure \ref{fig:BYDraFit} illustrates comparisons between our PTI
$V^2$ and archival RV data and our orbit model.  In Figure
\ref{fig:BYDraFit}a six consecutive nights of $K$-band visibility data
and visibility predictions from the best-fit model are shown, inset
with fit residuals along the bottom.  That there are only 1 -- 3 data
points in each of the nights follows from the brief time each night
that BY~Dra is simultaneously within both delay and zenith angle
limits.  Figure \ref{fig:BYDraFit}b gives the phased archival RV data
and model predictions, inset with a histogram of RV fit residuals.
The fit quality is consistent with previous PTI orbit analyses.

Spectroscopic orbit parameters (from VF79 and LM80) and our visual and
spectroscopic orbit parameters of the BY~Dra system are summarized in
Table \ref{tab:orbit}.  We give the results of separate fits to only
our $V^2$ data (our ``$V^2$-only Fit'' solution), and a simultaneous
fit to our $V^2$ data and the archival double-lined radial velocities
-- both with component diameters constrained as noted above.  All
uncertainties in parameters are quoted at the one sigma level.  We see
good statistical agreement between all the derived orbital parameters
with the exception of the LM80 period estimate.

\begin{table}
\begin{center}
\begin{small}
\begin{tabular}{ccc|cc}
\hline
Orbital			& VF79                & LM80   & \multicolumn{2}{c}{PTI 99} \\
\cline{4-5}
Parameter       	&                     &        & $V^2$-only Fit   & Full Fit \\
\hline \hline
Period (d)              & 5.9750998           & 5.975112 & {\em 5.975078}   & 5.975078       \\
                        & $\pm$ 8.7 $\times$ 10$^{-5}$ & $\pm$ 8.2 $\times$ 10$^{-6}$ &   & $\pm$ 1.0 $\times$ 10$^{-5}$ \\
T$_{0}$ (MJD)           & 41146.59            & 43794.193 & 51376.1407      & 51376.1232  \\
                        & $\pm$ 0.07          & $\pm$ 0.023 & $\pm$ 0.0087 &  $\pm$ 0.0045  \\
$e$                     & 0.36                & 0.3066  &   0.321            & 0.2978  \\
			& $\pm$ 0.03          & $\pm$ 0.0063 & $\pm$ 0.015   & $\pm$ 0.0032 \\
K$_1$ (km s$^{-1}$)     & 28.2 $\pm$ 1.0      & 28.55 $\pm$ 0.25 &                  & 28.31 $\pm$ 0.14  \\
K$_2$ (km s$^{-1}$)     & 28.8 $\pm$ 1.8      & 32.04 $\pm$ 0.35 &                  & 31.96 $\pm$ 0.15  \\
$\gamma$ (km s$^{-1}$)  & -24.47 $\pm$ 0.65   & -25.35 $\pm$ 0.14&                  & -25.431 $\pm$ 0.078 \\
$\omega_{1}$ (deg)      & 220 $\pm$ 5         & 229.3 $\pm$ 1.3   & 240.7 $\pm$ 8.2 & 232.21 $\pm$ 0.91 \\
$\Omega_{1}$ (deg)      &                     &                   & 106.7 $\pm$ 8.9 & 115.1 $\pm$ 1.3   \\
$i$ (deg)               &                     &                   & 153.0 $\pm$ 3.6 & 151.8 $\pm$ 3.5  \\
$a$ (mas)               &                     &                   & 4.376 $\pm$ 0.056 & 4.399 $\pm$ 0.050 \\
$\Delta K_{\rm CIT}$    &                     &                   & 0.69 $\pm$ 0.11 & 0.571 $\pm$ 0.061 \\
$\Delta H_{\rm CIT}$    &                     &                   & 1.39 $\pm$ 0.30 & 0.92 $\pm$ 0.24    \\
$\Delta V$              &                     & 1.15 $\pm$ 0.1    &                  &       \\
$\chi^2$/DOF	        &                     &                   & 1.1              & 2.4 (1.5 $V^2$/3.6 RV)  \\
$\overline{|R_{V^2}|}$  &                     &                   & 0.033 & 0.038 \\
$\overline{|R_{RV}|}$ (km s$^{-1}$)  &        & 0.55              &                  & 2.3 (0.49 COR)  \\
\hline
\end{tabular}
\end{small}
\caption{Orbital Parameters for BY~Dra.  Summarized here are the
apparent orbital parameters for the BY~Dra system as determined by
VF79, LM80, and PTI.  We give two separate fits to our data, with and
without including archival double-lined radial velocities in the fit.
Quantities given in italics are constrained to the listed values in
our model fits.  We have quoted the longitude of the ascending node
parameter ($\Omega$) as the angle between local East and the orbital
line of nodes measured positive in the direction of local North.  Due
to the degeneracy in our $V^2$ observable there is a 180$^\circ$
ambiguity in $\Omega$; by convention we quote it in the interval of
[0:180).  We quote mean absolute $V^2$ and RV residuals in the fits,
$\overline{|R_{V^2}|}$ and $\overline{|R_{RV}|}$ respectively.
\label{tab:orbit}}
\end{center}
\end{table}

\section{Comparisons With Hipparcos Model}
\label{sec:Hipparcos}
The Hipparcos catalog lists a circular photocentric orbital solution
for BY~Dra with a 114 day period (\cite{HIP97}), presumably in
addition to the well-established 6 d period A -- B motion.  As noted
above it is difficult to reconcile this hypothesis with the quality of
the existing short-period spectroscopic orbit solutions from VF79 and
LM80.  However if the A -- B system indeed did have a companion with
this period, unlike BY~Dra C it would lie within PTI's 1'' primary
beam, and if sufficiently luminous it would bias the visibility
measurements used in our BY~Dra A -- B visual orbit model.  We see no
indications of this in our orbital solutions; the quality of our
BY~Dra visual orbit solution is consistent with our results on other
systems.  But it remains possible we have misinterpreted our $V^2$
data in the binary star model fit, and we are motivated to consider
the 114-d periodicity hypothesis in the archival RV data.

We note that the Hipparcos model is a photocentric orbit, and
therefore calls for the A -- B system to exhibit a reflex motion with
radius $\geq$ 0.05 AU at the putative distance of BY~Dra.  The 114-d
orbit hypothesis is at high inclination (113$^{\circ}$), and therefore
would produce a radial velocity semi-amplitude for the A -- B system
barycenter $\geq$ 3.95 km s$^{-1}$.  This value is large compared to
fit residuals observed by VF79 and LM80 in their spectroscopic orbital
solutions (and by ourselves in the joint fit with our visibilities;
Table~\ref{tab:orbit}), suggesting the 114-d motion hypothesis is
unlikely.

\begin{figure}
\epsscale{0.8}
\plotone{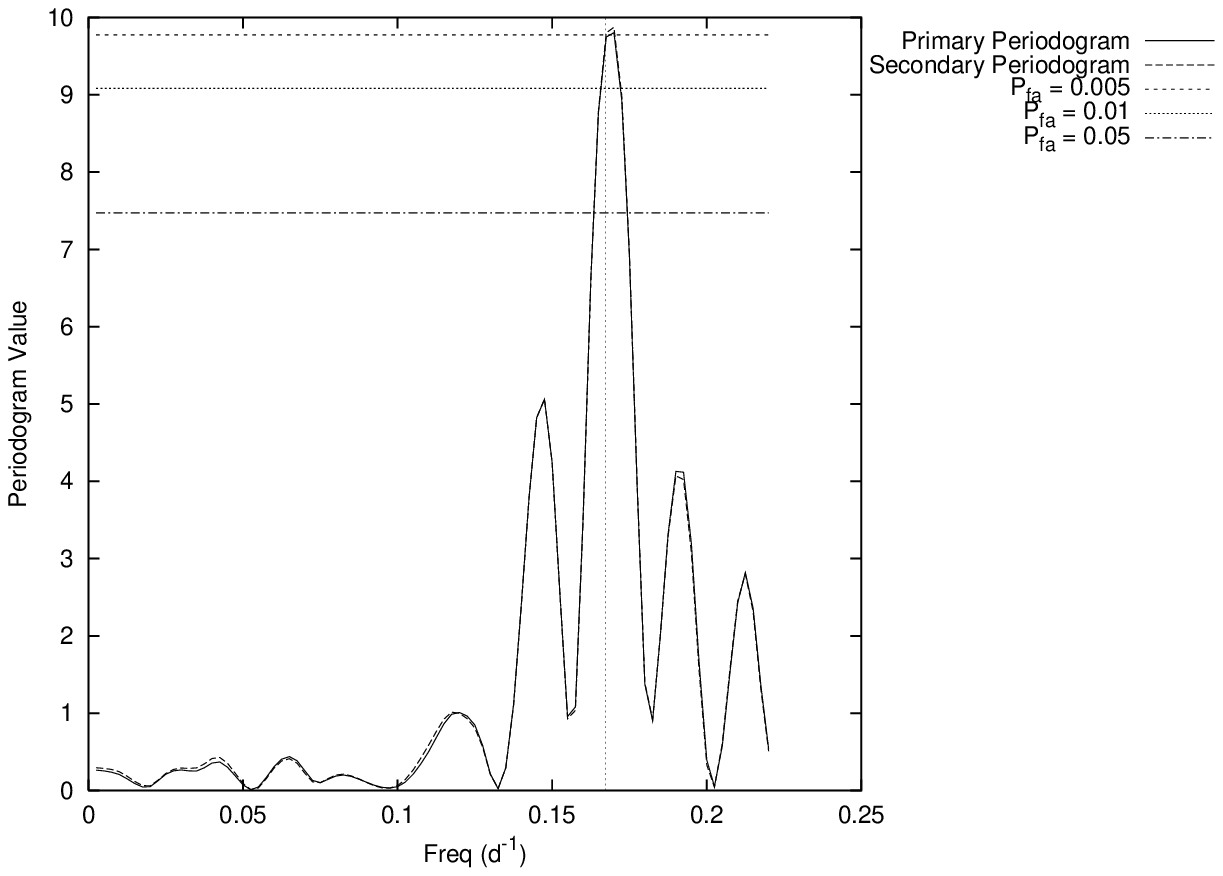}\\
\plotone{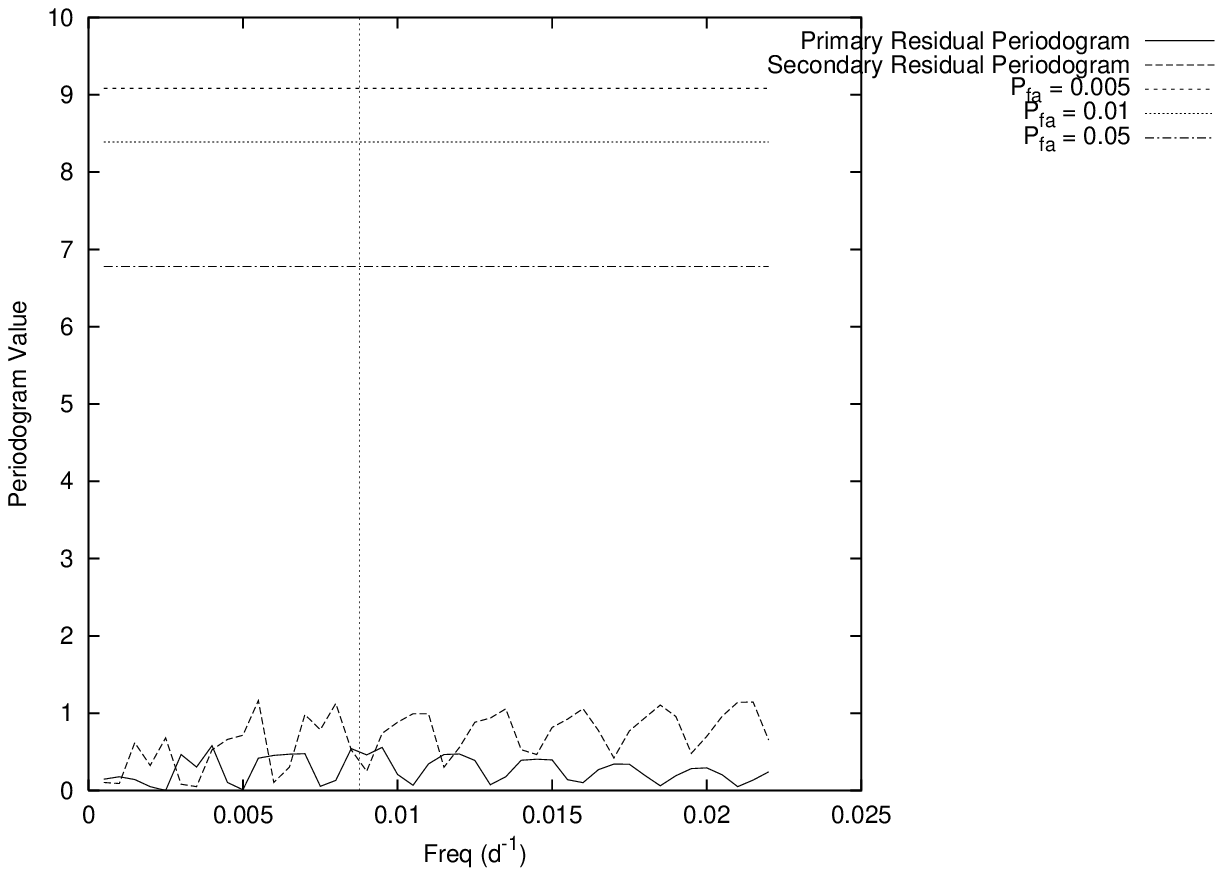}
\caption{Lomb -- Scargle Periodograms on LM80 BY~Dra Radial Velocity
Data.  To assess the 114-d orbit hypothesis for an additional
companion to the BY~Dra A -- B system we have performed Lomb --
Scargle periodogram analysis on double-lined BY~Dra data from LM80.
Top: standard periodograms on primary and secondary RV data from the
LM80 dataset.  Statistically significant response is seen at the
well-known 6 d A -- B period (vertical line).  Bottom: Low-frequency
periodogram of the residuals of the LM80 data to the best-fit A -- B
orbit model.  No significant response is evident at the putative 114 d
period (vertical line).
\label{fig:BYDraPeriodograms}}
\end{figure}

To quantify this issue we have considered Lomb-Scargle periodogram
analyses of the LM80 BY~Dra radial velocity data.  We have chosen to
use the LM80 data because it is the more precise sample, yielding an
rms residual of roughly 0.5 km s$^{-1}$ in our A -- B orbit analysis.
Figure \ref{fig:BYDraPeriodograms} gives periodograms of the LM80
primary and secondary radial velocity data.  First, Figure
\ref{fig:BYDraPeriodograms}a gives a periodogram of the primary and
secondary RV data, with probability of false alarm levels (P$_{fa}$)
as noted.  As expected, both component lines exhibit a significant
periodicity at the observed A -- B orbit frequency of approximately (6
d)$^{-1}$ (indicated by vertical line).  In the same plot we sample
the frequency of the 114-d orbit hypothesis, and no comparable
periodicity is evident on the scale of the A -- B motion.

Presuming the 114-d motion hypothesis might be superimposed on the 6-d
A -- B orbit, in Figure \ref{fig:BYDraPeriodograms}b we give a
periodogram of the LM80 primary and secondary RV fit residuals to the
6-d hypothesis fit.  In Figure \ref{fig:BYDraPeriodograms}b we have
adjusted to range of the periodogram to finely sample the frequency
range around the (114 d)$^{-1}$ hypothesis (indicated by vertical
line).  No significant periodicity is noted at this or any other
frequency.  The LM80 RV dataset spans roughly 400 days, so a 4 km
s$^{-1}$ amplitude periodicity in this dataset should have been
evident in our analysis.  Given these considerations, it seems
unlikely that our $V^2$ measurements and A -- B orbit model are
affected by the presence of a third luminous body.

\section{Discussion}
\label{sec:discussion}
The combination of the double-lined spectroscopic orbit and relative
visual orbit allow us to estimate the BY Dra A -- B component masses
and system distance.  However, because of the nearly face-on geometry
of the A -- B orbit, the accuracy of our inclination estimate, and
consequently the component mass estimates and system distance
estimate, is relatively poor.  The low-inclination geometry is
particularly difficult for astrometric studies because the astrometric
observable ($V^2$ in this case) becomes highly insensitive to small
changes in the inclination Euler angle.  Table \ref{tab:physics} lists
the physical parameters we estimate from our Full-Fit orbit solution.
The mass estimates are in a reasonable range for stars of this type
but are formally crude -- roughly 24\% 1-$\sigma$ uncertainties, as is
the system distance estimate -- roughly 8\% 1-$\sigma$.  In both cases
the dominant error is the inclination; to improve these mass and
distance estimates by a factor of two we must improve the inclination
estimate by a factor of two.  The orbital parallax with its large
error is in 1-$\sigma$ agreement with the Hipparcos triginometric
determination of 60.9 $\pm$ 0.75 mas, but we note that the Hipparcos
solution was derived jointly with the 3 mas, 114-d orbital hypothesis
that we believe to be suspect (\S \ref{sec:Hipparcos}).

\begin{table}
\begin{center}
\begin{small}
\begin{tabular}{ccc}
\hline
Physical	 & Primary           & Secondary \\
Parameter        & Component         & Component \\
\hline \hline
a (10$^{-2}$ AU) & 3.14 $\pm$ 0.36   & 3.55 $\pm$ 0.41   \\
Mass (M$_{\sun}$)& 0.59 $\pm$ 0.14   & 0.52 $\pm$ 0.13  \\
\cline{2-3}
System Distance (pc) & \multicolumn{2}{c}{15.2 $\pm$ 1.2 } \\
$\pi_{orb}$ (mas)    & \multicolumn{2}{c}{65.8 $\pm$ 5.4 } \\
\cline{2-3}
Model Diameter (mas) & {\em 0.60} ($\pm$ {\em 0.06}) & {\em 0.50} ($\pm$ {\em 0.05})   \\
M$_K$ (mag)      & 4.47 $\pm$ 0.18                   & 5.04 $\pm$ 0.18   \\
M$_H$ (mag)      & 4.48 $\pm$ 0.21                   & 5.40 $\pm$ 0.29   \\
M$_V$ (mag)      & 7.48 $\pm$ 0.18                   & 8.63 $\pm$ 0.18   \\
$V$-$K$ (mag)    & 3.019 $\pm$ 0.039                 & 3.598 $\pm$ 0.061  \\
\hline
\end{tabular}
\end{small}
\caption{Physical Parameters for BY~Dra.  Summarized here are the
physical parameters for the BY~Dra A -- B system as derived primarily
from the Full-Fit solution orbital parameters in Table
\ref{tab:orbit}.  Quantities listed in italics (i.e. the component
diameters, see text discussion) are constrained to the listed values
in our model fits.
\label{tab:physics}}
\end{center}
\end{table}

\begin{figure}
\epsscale{0.6}
\plotone{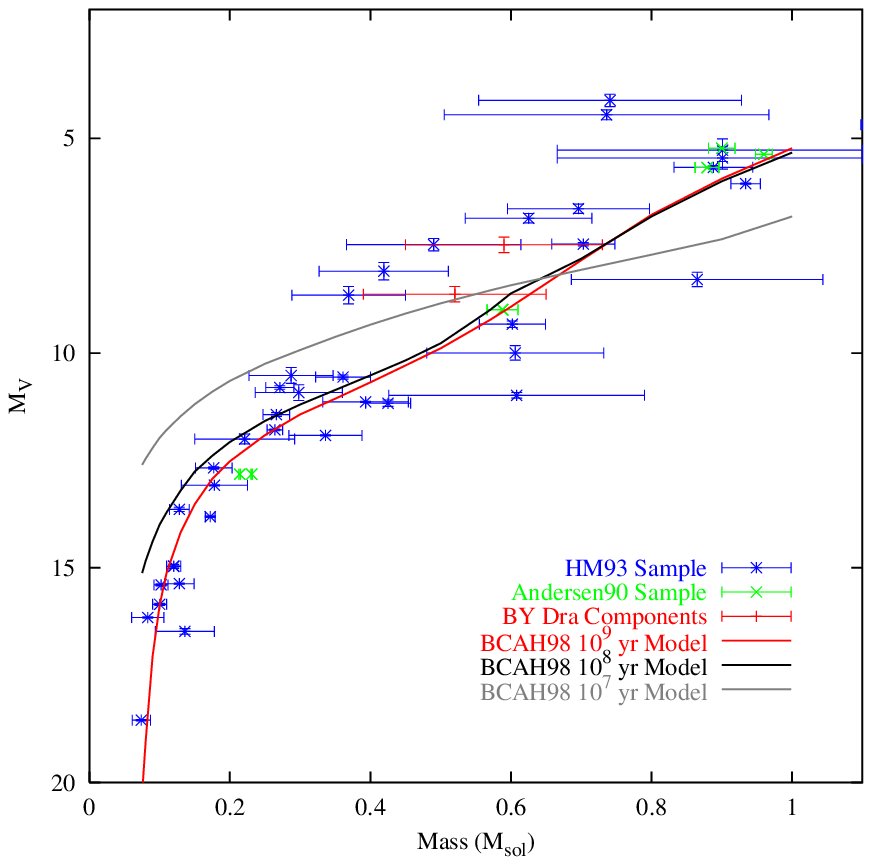}\\
\plotone{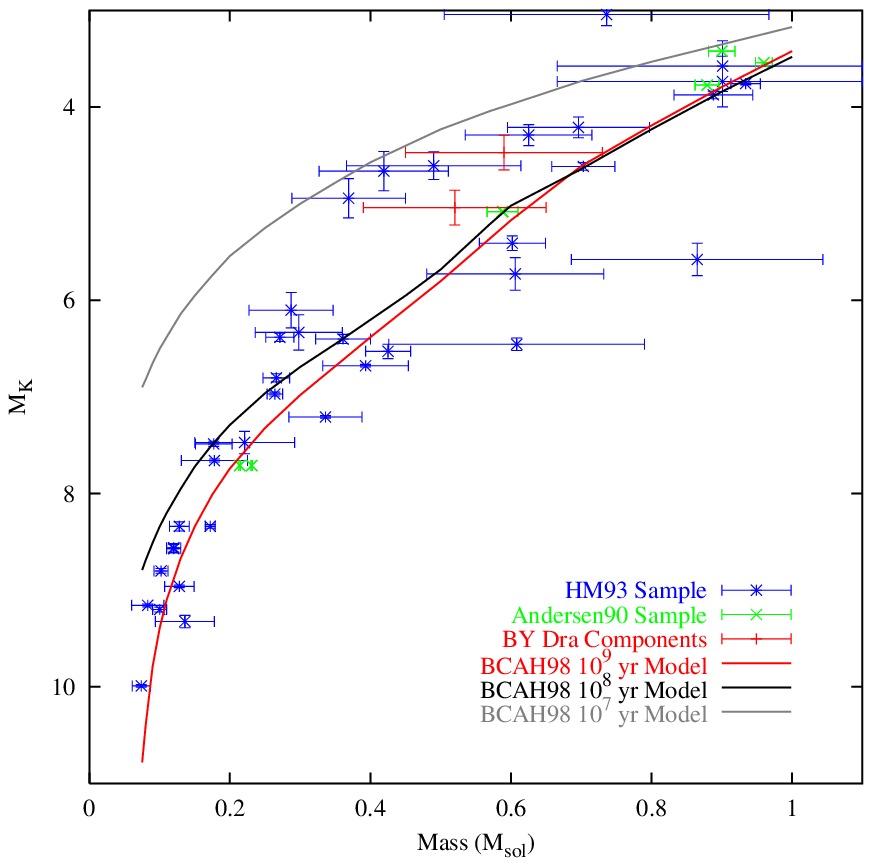}
\caption{BY Dra Components in Mass/Luminosity Space.  Here we give the
positions of the A and B BY Dra components in observable Mass/M$_V$
(upper pannel) and Mass/M$_K$ (lower pannel) spaces (an emulation of
Figure 3 from BCAH98), superimposing low mass objects from Henry \&
McCarthy (1993) and Andersen (1990) and BCAH98 solar-metalicity model
tracks at ages of 10$^9$, 10$^8$, and 10$^7$ yrs.  While our mass
determinations are currently too crude to be definitive, both
components appear over-luminous compared to the BCAH98 models, and it
is difficult to reconcile the deviations with age effects.
\label{fig:BYDraML}}
\end{figure}

Table \ref{tab:physics} also gives component absolute magnitudes and
$V$ - $K$ color indices derived from archival broad-band photometry,
our $H$ and $K$ component relative magnitudes, the LM80 $V$ relative
magnitude, and our system distance estimate.  The component color
indices are unaffected by errors in the system distance, and seem to
be in good agreement with the color indices expected from stars in
this mass range and the classical spectral typing of the BY~Dra
system.  The absolute magnitudes are somewhat uncertain from the
relatively poor distance estimate, but suggest a modest discrepancy
with the mass-luminosity models of Baraffe et al (1998, BCAH98).  As
depicted in Figure \ref{fig:BYDraML}, both components appear roughly
0.5 -- 1 mag brighter at $V$ and $K$ than the BCAH98 models evaluated
at our model masses would predict.  The cause for this discrepancy
could be abnormally large component sizes suggested by VF79, or the
component masses are underestimated in our analysis.  Clearly we need
to improve our orbit model so as to further constrain the BY Dra
component parameters.

To assess the VF79 component size/pre-main sequence hypothesis, the
most interesting measurement of the BY~Dra A -- B system would be
unequivocal measurements of the component diameters.  Unfortunately
our $V^2$ data are as yet insufficient to determine these diameters
independently.  Canonical sizes of 1.0 R$_{\sun}$ and 0.8 R$_{\sun}$
(indicated by model effective temperatures and our IR flux ratios)
yield angular diameters of approximately 0.6 and 0.5 mas for the A and
B components respectively.  At these sizes neither the $H$ nor
$K$-band fringe spacings of PTI sufficiently resolve the components to
independently determine component sizes.  Consequently we have
constrained our orbital solutions to these 0.6 and 0.5 mas model
values.  Our data does in fact prefer the slightly smaller primary
component diameter to the larger 1.2 R$_{\sun}$ size implied by VF79
and LM80 $v \sin i$ and rotational period measurements.  However
either primary model diameter is possible with the expected systematic
$V^2$ calibration errors.  Additional data we will collect in the
coming year (see below) may well place interesting upper limits on the
component sizes, but unambiguous resolution of the BY~Dra A and B
components will have to wait for a longer baseline infrared
interferometer; most likely the CHARA array currently under
construction on Mt.~Wilson
\footnote{{http://www.chara.gsu.edu/CHARAArray/chara\_array.html}}

The prospects for improving the BY~Dra orbital model in general, and
the orbital inclination estimate in particular are reasonable.  To
acheive 10\% mass determinations on the BY~Dra components we will need
to reduce the inclination uncertainty to approximately 1.5$^{\circ}$,
which implies a rough quadrupling of the PTI visibility data on
BY~Dra, or an improvement of our $V^2$ measurement precision by a
factor of two on a data set of similar size to that already collected.
Both are plausible; PTI is expected to be in normal operation for at
least another season, and hardware upgrades to the PTI fringe camera
are planned and should provide significant improvements in
sensitivity.

\acknowledgements The work described in this paper was performed at
the Infrared Processing and Analysis Center, California Institute of
Technology, and the Jet Propulsion Laboratory under contract with the
National Aeronautics and Space Administration.  Interferometer data
were obtained at Palomar Observatory using the NASA Palomar Testbed
Interferometer, supported by NASA contracts to the Jet Propulsion
Laboratory.  Science operations with PTI are conducted through the
efforts of the PTI Collaboration
(http://huey.jpl.nasa.gov/palomar/ptimembers.html), and we acknowledge
the invaluable contributions of our PTI colleagues.  We further
acknowledge I.N.~Reid (UPenn) for the compilation of mass-luminosity
data on Henry \& McCarthy and Andersen objects.

This research has made use of the Simbad database, operated at CDS,
Strasbourg, France.

\end{document}